# Low dimensional fragment-based descriptors for property predictions in inorganic materials with machine learning

Md Mohaiminul Islam*,

*Department of Mechanical Engineering, Temple University, Philadelphia, PA 19122, United States
*E-mail: md.mohaiminul.islam@temple.edu

## Abstract

In recent times, the use of machine learning in materials design and discovery has aided to accelerate the discovery of innovative materials with extraordinary properties, which otherwise would have been driven by a laborious and time-consuming trial-and-error process. In this study, a simple yet powerful fragment-based descriptor, Low Dimensional Fragment Descriptors (LDFD), is proposed to work in conjunction with machine learning models to predict important properties of a wide range of inorganic materials such as perovskite oxides, metal halide perovskites, alloys, semiconductor, and other materials system and can also be extended to work with interfaces. To predict properties, the generation of descriptors requires only the structural formula of the materials and, in presence of identical structure in the dataset, additional system properties as input. And the generation of descriptors involves few steps, encoding the formula in binary space and reduction of dimensionality, allowing easy implementation and prediction.

To evaluate descriptor performance, six known datasets with up to eight components were compared. The method was applied to properties such as band gaps of perovskites and semiconductors, lattice constant of magnetic alloys, bulk/shear modulus of superhard alloys, critical temperature of superconductors, formation enthalpy and energy above hull convex of perovskite oxides. An advanced python-based data mining tool matminer was utilized for the collection of data.

The prediction accuracies are equivalent to the quality of the training data and show comparable effectiveness as previous studies.

This method should be extendable to any inorganic material systems which can be subdivided into layers or crystal structures with more than one atom site, and with the progress of data mining the performance should get better with larger and unbiased datasets.

# 1. Introduction

Recently the material science community has experienced a growing interest in a data-driven predictive study with machine learning (ML). Previously, advancements in material science have been serendipitous and slow. With the advent of data-driven ML methods, the material science community has incorporated ML methods in their workflow and greatly benefitted from data-driven approaches in terms of computational cost and time. At the heart of ML models lies data. The performance of ML, as a researcher, improves with training and large datasets. As ML predictive models become more and more popular and robust, researchers emphasized collecting data to be used in ML. To solve the issue with data more and more databases are being developed, data are being collected from computational and experimental studies ML models have enjoyed great success in predicting properties of materials i.e., band gap[1–3], critical temperature of superconductor[4], shear, and bulk modulus[5], Debye temperature[6] and many more[7,8]. One of the major advantages of ML models is that the prediction takes place in a very short time. Standard material characterization practices such as calculating the band structure are known to be notorious with Density functional theory (DFT) where calculations can take several days for prediction and going beyond standard DFT with random phase approximation or GW approximation[9], the calculations can become so expensive computationally that it can become entirely impractical. Whereas ML methods offer very fast prediction with comparable accuracy with the help of previously calculated properties from data, as a result, the ML method has the ability to offer a fast screening of materials for targeted properties.

The performance of ML models depends on the choice of descriptors as well as the quality of data. But the choice of descriptors is not exclusive nor unique for a system and the data is often biased. To circumvent this issue of lack of data, several databases have been created for ML models i.e., Citrine Informatics[10], Materials Project (MP)[11], Materials Data Facility (MDF)[12], and Materials Platform for Data Science (MPDS)[13]. Nonetheless, most of the open-source databases suffer from a lack of experimental data and often are seen to be biased toward favorable outcomes and the datasets sometimes lack uniformity.

Developing descriptors for materials system holds paramount importance for the performance of ML models. There has been a considerable effort in developing universal descriptors. Several representations are available such as: Coulomb matrix[14], which contains information on atomic nuclear repulsion and the potential energy of free atoms, graph descriptions[15], in the solid state, representations based on radial distribution functions[16], Voronoi tessellations[17] or property-labelled materials fragments[6] to name a few. quantitative structure-activity relationship modelling coupled with virtual screening of chemical libraries have been largely successful in the discovery of novel bioactive compounds[18]. However, several descriptors suffer from "curse of dimensionality" due to high dimensional features and require very large datasets to predict properties with reasonable accuracy. And every descriptor doesn't work with all properties.

In this study, a new fragment-based descriptor of materials has been proposed. To assess the validity of the descriptors on different materials system and different properties eight different datasets were chosen to account for different material space i.e., semiconductors, perovskites, metal, or non-metal, binary, tertiary, or quaternary systems, and the nonuniformity of datasets and to show the capability and performance of the proposed model in aforementioned systems. First, the method for generating descriptors has been described, then implemented along with ML models. Next, the effectiveness of the proposed method is assessed by applying the proposed approach to five established datasets and comparing eight predicted properties i.e., band gap of double perovskite, critical temperature of superconductivity, lattice constant and band gap of metal halide perovskites, experimental band gap of semiconductors, anisotropy, shear, and bulk modulus from superhard alloys dataset. To proof its applicability this model has been applied to two new datasets previously not used for property prediction to our knowledge, namely the Heusler magnetic dataset to predict lattice constant and $ABO_3$ perovskite dataset for prediction of formation energy and energy above hull convex. Then the model has been extended to work with interfaces which are demonstrated on the interface thermal conductivity (ITR) dataset. This extension also demonstrates that the proposed descriptor can be easily extended to work with any material systems with layers or crystal structure with more than one atom sites.

A comparative study has been shown with previous studies to exhibit applicability and performance of the presented method. Several other investigations have predicted a subset of properties presented in this study, such as Miyazaki et. al.[19] have predicted properties of half Heusler alloys while in this study full, inverse, and half Heusler alloys have been included.

## 2. Methods

Descriptors are of paramount importance for property prediction with ML algorithms. Several ML study have demonstrated that the choice of descriptors have bigger influence on accuracy of prediction than the choice of ML algorithms when predicting material property[20,21]. Therefore, proper development of descriptor is imperative for ML success. Earlier studies have shown that the fragment-based descriptor perform well in polymers, organic molecules and drug design[22,23]. For present work, fragment-based descriptor was incorporated as a starting point. In fragment-based descriptors, material systems are represented as numeric value understandable to ML models. The fragment-based descriptor for the presented work starts by dividing the formula of the material system into their components. Any molecular invariant (i.e., same chemical formula with different material property) can be handled by additional condition such as temperature, Laue group, space group etc.

Figure 1 shows the scheme for constructing LDFD. Given a formula of a material system, the first step is to separate the formula into associated components. For example: $V_2CrSi$ with $D022$ structure is a full Heusler alloy, to predict its property, it is decomposed into constituent components $V, Cr$, and $Si$. There is another variation of $V_2CrSi$ with $Xa$ structure belonging to inverse Heusler alloy group. In these cases where chemical structure is same shown in Figure 1(b), the structure type is added as a descriptor to differentiate between these alloys. If the chemical formula contains larger groups or ions, then the formula is simplified for ML models. For example: $(CH_3NH_3)PbI_3$ is a metal halide perovskite, the formula can be simplified (shown in Figure 1(a)) as $MPbI_3$ where $M = CH_3NH_3$. In general, the decomposition of the formula is simple to implement. The position of the atom/compounds carries momentous importance especially for layered materials as it can dictate material property. For a given spatial arrangement of chemical elements, the distribution of electrons and a wide range of physical responses can be described. The second step is to encode components into binary values. In the proposed method, this is done for each position of atoms/groups to keep the spatial information intact. The advantage of encoding by position is it can capture positional importance of the atoms/groups and it also and makes it easier to capture the whole material system without assigning importance to any particular atoms/compounds.

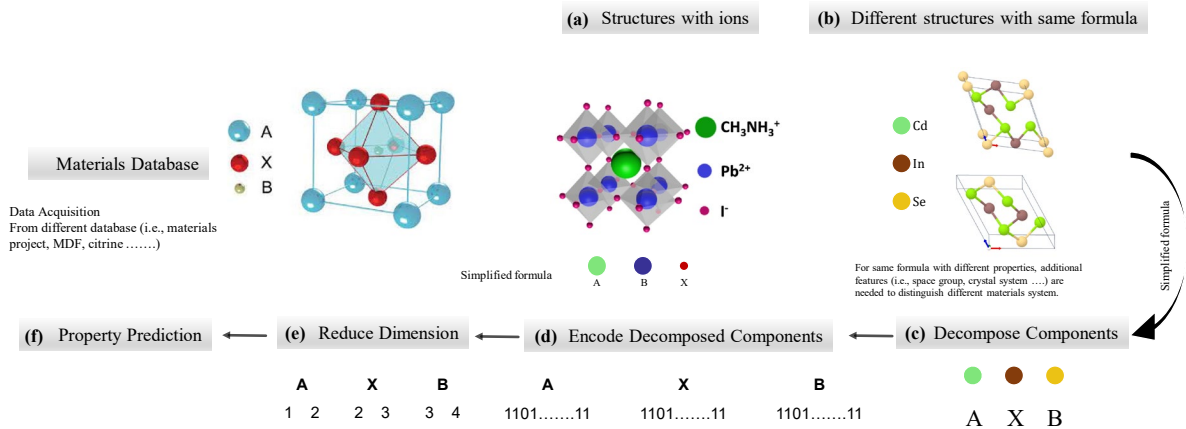

Figure 1: Schematic description for the construction of the Low Dimensional Fragment Descriptor (LDFD). First, data is acquired from material databases and their formula is simplified if necessary then the formula is decomposed to be embedded. Property is predicted with a lower dimensional descriptor space of the encoded descriptor.

The final step is to reduce dimensionality. Position based binary on the decomposed formula results in discrete values. Classical methods of dimensionality reduction handle continuous data.

The proposed descriptor has several advantages over other types of chemical or physical descriptors [24], including simplicity of calculation, storage and lower dimension. These advantages should speed up the prediction process. However, there are a few disadvantages. Models built with fragment descriptors perform poorly when presented with new fragments for which they were not trained. And the model needs to be trained each time new materials are included in the databases. However, with modern advancements with material informatics more and more comprehensive databases are constructed, such efforts will surely help fragment-based descriptors overcome that disadvantage.

Mindful of these constraints, the proposed fragment-based descriptors have been shown to work with different materials systems from semiconductors to different perovskites, magnetic alloys to superhard alloys, and superconducting materials. Due to low dimension this descriptor also performs reasonably well with smaller datasets such as: the metal halide perovskite dataset contains only 873 data. Larger datasets are known to aid all ML models, thus, with more and more data collection ML models should perform better. And the proposed model can be also extended to work with interfaces as well, details are described in section 7.

## 3. Dimensionality Reduction

High dimensional descriptor space requires proportional amount of data for training to obtain reasonable accuracy in prediction. For such cases, Dimensional reduction techniques have been developed to reduce dimensionality of the original high-dimensional data into lower-dimensional data without significant loss of information. With lower dimensional descriptors, reasonable accuracy in prediction can be achieved with much smaller amount of data. There are several choices for dimensionality reduction, among them principal component analysis (PCA) is by far the most common and successful algorithm for data compression, visualization, and feature discovery. PCA calculates the linear projections of the data with maximum variance, or map data to such a lower dimensional subspace that implicitly minimizes the reconstruction error under the squared error loss. Classical PCA methods have been developed for numerical data not binary data. Orthogonal projection of high-dimensional data $x$ on low-dimensional space spanned by $B$ of rank $R$ such that $B(J \times R)$, $R \ll \min(I,J)$ can be expressed as:

$$\min(\mu, B) \quad \sum_{i}^{I}(x_i - \mu - BB^T(x_i - \mu))^2$$
$$\text{subject to} \quad B^T B = I_R \tag{1}$$

$\mu$ is column offset term, $x_i$, $i = 1,2 \ldots l$, the $i$th row of matrix $x$, is the $J$-dimensional measurement of the $i$th object and $I_R$ is $R$-dimensional identity matrix. The classical PCA was derived from a geometrical perspective. Bishop et al. [25] have derived PCA from a probabilistic perspective, called probabilistic PCA. The conditional distribution of $x_i$ which is regarded as noisy observation of true data $\theta_i$ in a low dimensional space with a normal distribution with zero (0) mean and constant variance $\sigma^2$. The variables can be approximated as $x_i = \theta_i + \varepsilon_i$, and $\theta_i = \mu + Ba_i$, $\mu$ is the offset term as before; $B$ contains the coefficients; $a_i$ represents the low-dimensional score vector. The noise term $\varepsilon_i$ is assumed to follow a multivariate normal distribution with zero (0) mean and constant variance $\sigma^2$, $\varepsilon_i \sim N(0, \sigma^2 I_J)$. The maximum likelihood estimation of the distribution of $x_i$ with mean $\theta_i$ and constant variance can be expressed as:

$$\max(\mu, a_i, B) \sum_{i}^{l} \log((p(x_i|\mu, a_i, B))$$

$$= \sum_{i}^{l} \log(N(x_i|\mu + Ba_i, \sigma^2 I_j))$$

$$= \sum_{i}^{I} \sum_{j}^{J} \log\left(N(x_{ij}|\mu + a_i^T b_j, \sigma^2)\right) \quad (2)$$

$a_i$ represents the low-dimensional score vector, $x_i|\mu$ represents conditional distribution of $x_i$ for a given $\mu$, and $b_j$ is the $j$th entry of $B$ matrix. As PCA is based upon Gaussian assumption, it is only appropriate for continuous numerical valued data. Considerable effort has been given to extend PCA to work with binary data [26,27]. One such extension is logistic PCA[26], which is a similar extension to classical PCA compared to logistic linear regression to linear regression.

## 4. Logistic PCA

To extend classical PCA to logistic PCA, instead of using gaussian distribution, binary distribution is used. Logistic PCA is based on a multivariate generalization of the Bernoulli distribution [26]. The Bernoulli distribution for a univariate binary random variable $x \in \{0, 1\}$ with mean $p$ is given by:

$$P(x|p) = p^x (1-p)^x \quad (3)$$

It's important to distinguish that $x$ used in Eqn.(3) is binary whereas in previous section $x$ was used to denote continuous variable. Eqn.(3) can be re-written in terms of the log-odds parameter $\theta = \log\left(\frac{p}{1-p}\right)$ and logistic function $\sigma(\theta) = (1 + e^{-\theta})^{-1}$. The Bernoulli distribution can be given in new variables by:

$$P(x|\theta) = \sigma(\theta)^x \sigma(-\theta)^{1-x} \tag{4}$$

The log-odds and logistic function are, respectively, the natural parameter and canonical link function of the Bernoulli distribution expressed as a member of the exponential family. A simple multivariate generalization yields the logistic PCA model. If $X_{nd}$ denotes the elements of an $N \times D$ binary matrix, where $N$ rows capture the observation vector of a $D$-dimensional binary space. The probability distribution can be expressed as:

$$P(X|\Theta) = \Pi_{nd}\, \sigma(\Theta_{nd})^{X_{nd}}\, \sigma(-\Theta_{nd})^{1-x_{nd}} \tag{5}$$

where $\Theta_{nd}$ denotes the log-odds of the binary random variable $X_{nd}$. The log-likelihood ($\mathcal{L}$) of binary data under this model is given by:

$$\mathcal{L} = \sum_{nd} [x_{nd} \log(\sigma(\Theta_{nd})) + (1 - x_{nd})\log(\sigma(-\Theta_{nd}))] \tag{6}$$

By maximizing the log-likelihood, low dimensional structure in the data can be discovered. By constraining $\Theta$ to occupy a lower dimensional subspace, compact representation similar to classical PCA can be obtained, with $L \ll D$ where $L$ is the dimensionality of the latent subspace and $D$ is dimensionality of binary data. The log-odds matrix $\Theta$ can be parameterized in terms of two smaller matrices $U$ and $V$ with a bias vector $\Delta$. In terms of these parameters, the $N \times D$ matrix $\Theta$ is represented as:

$$\Theta_{nd} = \sum_{l} U_{nl} V_{ld} + \Delta_d \tag{7}$$

where $U$ is an equally tall but narrower matrix with dimension $N \times L$, $V$ is a shorter but equally wide matrix with dimension $L \times D$, $\Delta$ is a $D$-dimensional vector, and the sum over the subscript $l$ in Eqn.(7) makes explicit the matrix multiplication of $U$ and $V$. The parameters $U, V$ and $\Delta$ in this model play roles similar to the linear coefficients, basis vectors, and empirical mean computed by classical PCA for continuous data. Though the bias vector $\Delta$ in this model could be absorbed by a redefinition of $U$ and $V$, its presence permits a more straightforward comparison to linear PCA of mean-centered data.

Logistic PCA can be applied to binary data in largely the same way that classical (or linear) PCA is applied to continuous data. Given binary data $x$, the parameters $U$ and $V$ are computed, then $\Delta$ that maximizes (at least locally) the log-likelihood in Eqn.(6). An iterative least square method is used for maximizing Eqn.(6). Thus, having estimated these parameters from training data $x$, a low dimensional representation $U'$ can be computed of previously unseen (or test) data $x'$ by locating the global maximum of the corresponding test log-likelihood $\mathcal{L}'$ (with fixed $V$ and $\Delta$). Logistic and linear PCA can both be viewed as special cases of the generalized framework described by Collins et al [26]. This is done by writing the log-likelihood in Eqn.(6) in a more general form:

$$\mathcal{L} = \sum_{nd} \Theta_{nd} x_{nd} - G(\Theta_{nd}) + log(P_o(x_{nd})) \qquad (8)$$

In this more general formulation, the function $G(\Theta_{nd})$ in Eqn. (8) is given by the integral of the distribution's canonical link, while the term $P_o(x_{nd})$ provides a normalization but has no dependence on the distribution's natural parameter. Note that logistic PCA has some of the same short comings as linear PCA particularly, it does not define a proper generative model that can be used to handle missing data or infer a conditional distribution $P[U|x]$ over the coefficients $U$ given data $x$.

Two principal components were used for the construction of LDFD and Random Forest regressor was used for prediction.

## 5. Datasets and Predictions

With the advent of new computing power more and more databases are being generated and made public. In this work, a data mining toolkit matminer have been used to retrieve four of the datasets from open-source databases such as Materials Project (MP), Citrine Informatics, The Materials Data Facility (MDF), The Materials Platform for Data Science and many more. And three of the datasets have been collected from other sources [2–4].

The double perovskite (represented by the chemical formula $AA'BB'O_6$) dataset contains 53 stable cubic perovskite oxides which were found to have a finite bandgap in a previous screening based on single perovskites [28]. These 53 parent single perovskites contained fourteen different A-site cations ($Ag, Ba, Ca, Cs, K, La, Li, Mg, Na, Pb, Rb, Sr, Tl$ and $Y$) and ten B-site cations ($Al, Hf, Nb, Sb, Sc, Si, Ta, Ti, V, Zr$). Four cations ($Ga, Ge, In$ and $Sn$) were found to appear on either A- or B-sites. A total of 1378 unique double perovskites are possible of which 72 double perovskites are metallic (or have a very small bandgap $< 0.1\ eV$) and are not included in the database which results in 1306 unique double perovskites. The reported bandgaps are computed using density functional theory (DFT) [29] as implemented in the GPAW code [30] with the Gritsenko, van Leeuwen, van Lenthe and Baerends potential (GLLB) [31], further optimized for solids by Kuisma et al. [32]. The double perovskites also show structural symmetry (meaning the structure is invariant with respect to swapping of the two cations at $A$-site and $B$-site, i.e., $AA'BB'O_6$, $A'ABB'O_6$ and $AA'B'BO_6$ are all identical systems). So, number of possible compounds are 3918. The low dimensional fragment-based descriptor used in this present work uses 8-dimensional descriptor for prediction of band gap. The dataset contains minimum band gap of $0.1\ eV$ and maximum of $8.34\ eV$ with standard daviation $1.58\ eV$. 10-fold-cross validation gives a mean coefficient of determination $R^2$ of 0.91. The quantification of error was done with $MAPE$ (mean absolute percentage error) and $RMSE$ (root mean squared error). For double perovskite dataset LDFD gives $MAPE$ of 0.71% and $RMSE$ of $0.62\ eV$. Compared to the original work ($R^2 = 0.94$ and $RMSE = 0.78\ eV$), the predictions are comparable with half of the descriptor dimension. Low dimensions should facilitate speeding of the prediction process.

Superconductivity dataset was collected form Stanev et.al. [4], they extracted data from the SuperCon database [33]. From the database, they extracted ~16,400 compounds, of which 4000 have no thermal conductivity reported which were excluded from the dataset. Roughly, 5700 compounds are cuprates and 1500 are iron based. The remaining set of about 8000 is a mix of various materials, including conventional phonon-driven superconductors. This dataset contains wide variety of compounds. As the LDFD depends on number of components within the compounds, this dataset demonstrates performance of LDFD with different component system in the dataset. This dataset contains compounds with minimum 2 and up to 8 components. Figure 2(a) shows description of the dataset based on presence of different component in a material. Due to this variation in number of components LDFD requires 16 dimensional descriptor compared to the original work which requires 10 dimensional descriptors. Prediction of critical temperature with LDFD with 10 fold cross validation gives $R^2$ of 0.92 and $MAPE$ of 0.95 and $RMSE$ of $12.90\ K$. Compared to the original work $R^2$ of 0.88. This dataset demonstrates the applicability of LDFD with variable component system provided enough examples (this dataset contains more than 12000 data).

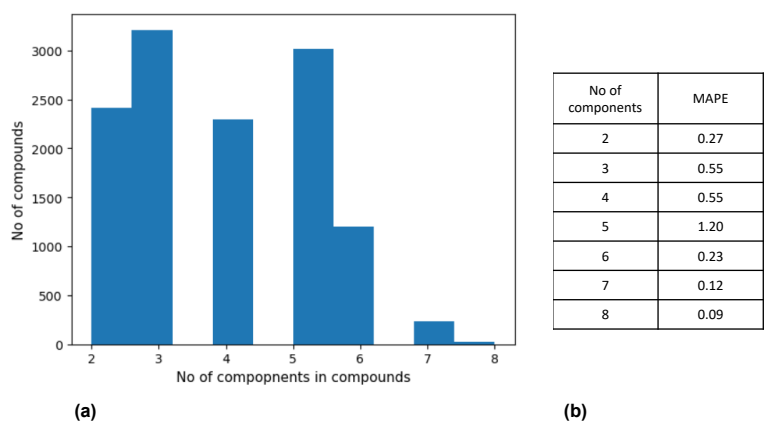

(a) (b)

Figure 2: Component based error analysis of Superconductivity dataset. (a) Shows number of components in the superconductivity dataset. (b) Error distribution for different component system.

The error distribution in Figure 2(b) shows no particular inclination towards any component. This is reassuring for LDFD to perform in a dataset containing different component materials.

The metal halide perovskites (MHPs) dataset was taken from Saidi *et. al.* [28] They performed high-throughput computational screening of $ABX_3$ MHPs generated from the parent MHP $MAPbI_3$ ($MA = CH_3NH_3$) structure, variations were made only at the "A-ion" site. At the A site, $Cs$ in addition with 18 different organic molecules are considered namely: ammonium $[NH_4]^+$, $[AsH_4]^+$, $[PH_4]^+$, $[AsH_4]^+$, $[PF_4]^+$, methylammonium $[CH_3NH_3]^+$, $[CH_3PH_3]^+$, $[CH_3AsH_3]^+$, hydrazinium $[(H_3N)(NH_2)]^+$, azetidinium $[(CH_2)_3NH_2]^+$, formamidinium $[NH_2(CH)NH_2]^+$, $[NH_2(CH)PH_2]^+$, $[NH_2(CH)AsH_2]^+$, imidazolium $[C_3N_2H_5]^+$, dimethylammonium $[(CH_3)_2NH_2]^+$, acetamidinium $[NH_2 - C - CH_3 - NH_2]^+$, ethyl ammonium $[(C_2H_5)NH_3]^+$, and hydroxylammonium $[H_3NOH]^+$. In total, the dataset comprises of just 862 total compounds. The bandgap of the MHPs in the dataset are computed using DFT. For the purpose of machine learning, the formula of the compounds was simplified (i.e., $E = [(C_2H_5)NH_3]^+$, $Am = [NH_4]^+$ and so on). Among the dataset this is the smallest which causes the $R^2$ to be lowest and $MAPE$ to be the highest among all the example datasets. However, it certifies that LDFD can handle predict material properties for complex formula. There are two target properties to predict band gap and lattice constant, 10 fold cross validation gives $R^2$ of 0.72 for band gap and 0.78 for lattice constant. Percentage error is just over 1 for band gap with $MAPE$ 1.07 and for lattice constant 1.71. And $RMSE$ for bandgap is 0.5 $eV$ and for lattice constant 0.12 $A^o$. The original work reports $RMSE$ of 0.14 $eV$ for band gap and 0.16 $A^o$ for lattice constant. Direct comparison is difficult as they used convolutional neural network (CNN) and hierarchical CNN (HCNN) for prediction of the properties which requires visual data. Though there are differences in prediction methods the errors are comparable. LDFD should perform better with larger datasets.

The Semiconductor dataset was originally collected from Brgoch *et.al.* [1] with the help of data mining toolkit matminer[34]. The dataset contains experimentally measured bandgaps of 6354 inorganic semiconductors of which more than 4000 are unique. After keeping semiconductors with band gap $> 0\ eV$ the total number of semiconductors goes down to 3434. This dataset also contains compounds with variation in component number (2 components system to 5 components system). After superconductivity dataset this dataset should provide additional evidence for LDFD to be able to predict properties from a dataset with variation in component number with much lower descriptor dimension. In error analysis 5 component systems were discarded as there were only a few data in the dataset.

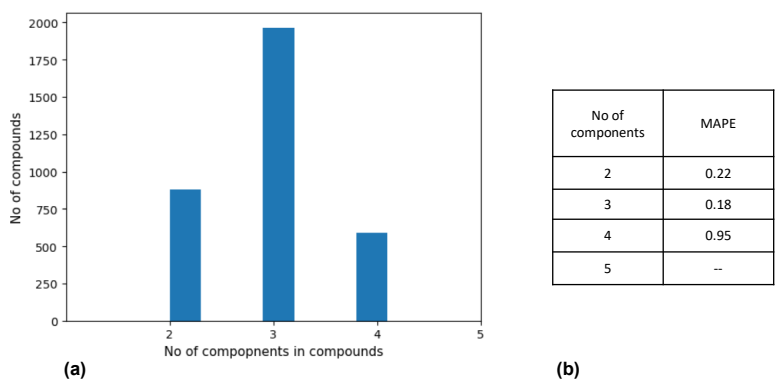

Figure 3: Component based error analysis of Semiconductor band gap dataset. (a) Shows number of components in the Semiconductor band gap dataset. (b) Error distribution for different component system.

The superhard materials dataset was collected with the help of matminer, this dataset was originally generated from Brogch *et.al.* [5], it contains 2574 Ultra incompressible, Superhard Materials. Reported shear and bulk modulus were found from experiments. 20 fold cross validation shows $R^2$ of 0.95 for bulk modulus and 0.94 for shear modulus in training dataset. Compared to original work $R^2$ of 0.94 for bulk modulus and 0.88 for shear modulus and reported $RMSE$ of 17.2 $GPa$ for bulk modulus and 16.5 $GPa$ for shear modulus. LDFD shows testing $RMSE$ of 6.89 $GPa$ and 2.94 $GPa$ for bulk and shear modulus respectively. But, little lower $R^2$ value of 0.85 and 0.75 for bulk and shear modulus respectively. Lower value of $R^2$ for shear modulus can be attributed how the shear modulus values are distributed.

The distribution is very skewed towards for shear modulus $< 100\ GPa$. Larger dataset with uniform distribution of the shear modulus should result in better prediction.

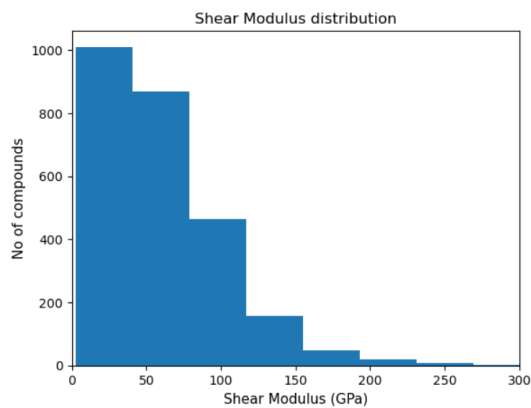

Figure 4: Shear modulus distribution in the dataset. clearly, there are inadequate data for shear modulus > 75 for reliable prediction.

They have also reported similar issue overestimation of low values ($< 75\ GPa$) and a slight underestimation at high values ($> 250\ GPa$).

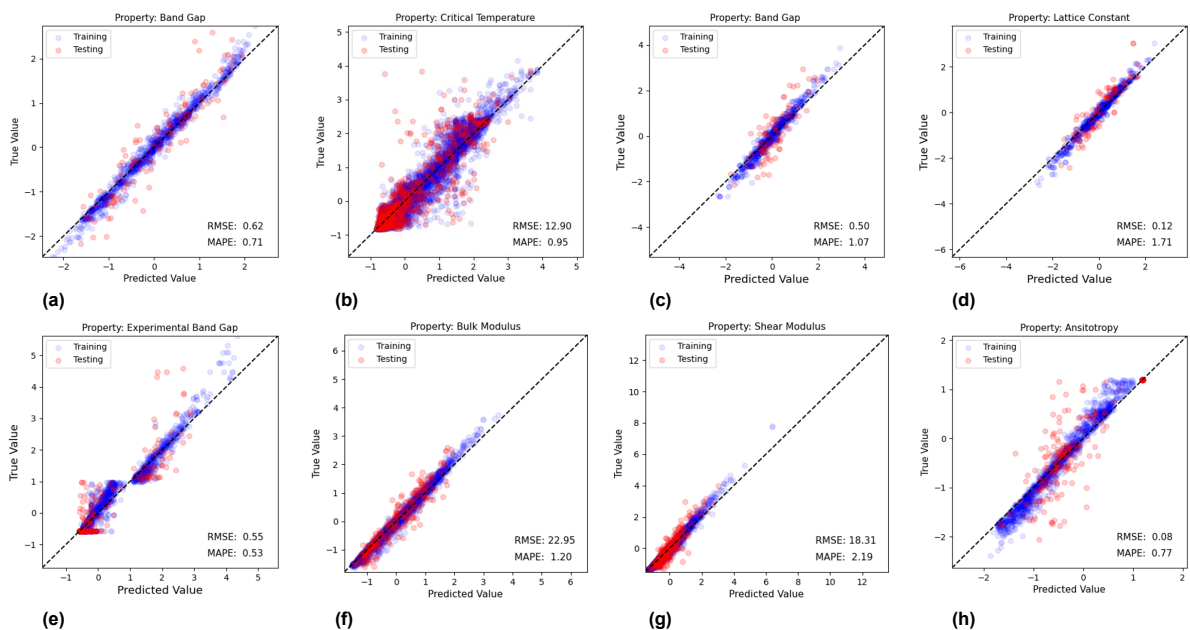

Figure 5: Plots for the eight ML predicted true values for the regression model (a) Band gap energy predicted from Double perovskite dataset (b) Critical temperature predicted from superconductivity dataset. (c-d) Band gap energy and lattice constant predicted from metal halide perovskite dataset. (e) Experimental

band gap predicted from semiconductor band gap dataset. (f-h) Bulk modulus, shear modulus and anisotropy predicted from super hard alloy dataset.

Figure 5 and Table 1 summarizes the results of the 10-fold cross validation analysis for the eight different regressions. Only the metal halide perovskite dataset has $R^2$ value lower than 0.8 due to small amount of data. For comparison with the original work super hard alloy dataset reported values are training values. But, in Figure 1(f-h) testing values with 90 to 10 splitting has been shown.

| Dataset | Properties | Criterion (Present Work) | | | Descriptor Dimension | | Comparison | | |
|---|---|---|---|---|---|---|---|---|---|
| | | $R^2$ | MAPE | RMSE | This work | Ref. | $R^2$ | RMSE | Ref. |
| Double Perovskite | Band Gap | 0.91 | 0.71 | 0.62 | 8 | 16 | 0.94 | 0.78 | 2 |
| Superconductivity | Critical Temperature | 0.92 | 0.95 | 12.90 | 16 | 10 | 0.88 | - | 4 |
| Metal Halide Perovskite | Band Gap | 0.72 | 1.07 | 0.50 | 10 | 11 | - | 0.14 | 3 |
| | Lattice Constant | 0.78 | 1.71 | 0.12 | | | | 0.16 | |
| Semiconductor band gap | Experimental Band Gap | 0.90 | 0.53 | 0.55 | 10 | 136 | 0.90 | 0.45 | 1 |
| Super hard alloys | Bulk Modulus | 0.95 | 0.56 | 6.89 | 19 | 150 | 0.97 | 14.3 | 5 |
| | Shear Modulus | 0.94 | 0.88 | 2.94 | | | 0.88 | 18.4 | |
| | Anisotropy | 0.88 | 0.77 | 0.08 | | | - | - | |

Table 1: Statistical summary of the predictions for eight properties collected from five different dataset with comparison to the original work.

## 6. Model validation

Although the performances of the ML models can be estimated from cross validation. But there is no better way to validate but with new datasets and predict new properties. To emphasize on this, two new datasets (to our knowledge, no study has been published to predict properties with these datasets) were used which to our knowledge first time being used for prediction. One is Heusler magnetic alloy dataset [35] and another $ABO_3$ perovskite dataset [36]. Both datasets are available through matminer.

The Heusler alloys, named in honor of their discoverer Dr. Heusler are intermetallic remarkable for the fact that, in certain proportions, they are ferromagnetic, although the component metals are not ferro-magnetic. Due to their tunable semiconducting properties these compounds are currently being heavily investigated for sustainable technologies such as solar energy and thermoelectric conversion. There are two families of Heusler compounds: half-Heusler compounds $ABC$, and (full-Heusler compounds $A_2BC$. The components are metals, where typically $A$ is a large electropositive metal, $B$ is a transition metal, and $C$ is an electronegative metal (usually a p-block metalloid). There is a complication: inverse Heusler compounds $ABAC$ are formed with the content of $A$ doubled and that of $B$ halved relative to the normal Heusler compounds $AB_2C$. The inverse Heusler structure consists of four sites within a face-centered cubic lattice and has lower symmetry. The dataset consists of 1153 alloys of which 576 are full, 449 are half and 128 are inverse Heusler alloys. Lattice constant of these alloys have been calculated from DFT. 10 fold cross validation gives a mean of $R^2$ around 0.95 and $RMSE$ and $MAPE$ of 0.31 and 0.48 respectively.

Because there is duplicate formula with different structure and different property, structure type of the alloys has been used as a descriptor in conjunction with the formula. So, the descriptor dimension became 11.

Second dataset used for validation is the $ABO_3$ perovskite dataset [36], collected from matminer. $ABO_3$ perovskites are oxide materials that are used for a variety of applications such as solid oxide fuel cells, piezo-, ferro-electricity and water splitting. Due to their remarkable stability with respect to cation substitution, new compounds for such applications potentially await discovery.

The ideal $ABO_3$ cubic perovskite crystal structure is composed of a $B$ cation that is octahedrally 6-fold coordinated with oxygen atoms and an $A$ cation that is 12-fold coordinated by oxygen atoms.

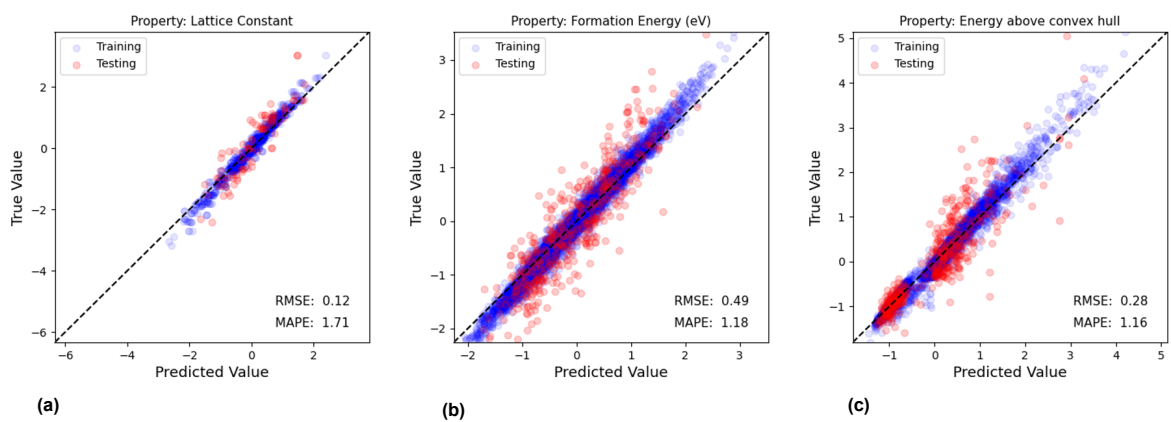

Figure 6: Plots of the new datasets (a) Lattice constant predicted from Heusler magnetic alloy dataset (b-c) Formation energy and Energy above convex hull predicted from $ABO_3$ perovskite dataset.

The dataset contains 4536 perovskite compounds of the form $ABO_3$. For ML purposes data with energy above convex hull as discarded, resulting in 4047 $ABO_3$ perovskites. Lowest distortion of the compounds was added with the formula as descriptors and final dimension of descriptor became 6. Summary of $R^2$, $MAPE$ and $RMSE$ have been recorded in Table 2.

| Dataset | Properties | $R^2$ | MAPE | RMSE |
|---|---|---|---|---|
| Heusler Magnetic Alloy | Lattice Constant | 0.95 | 0.48 | 0.31 |
| $ABO_3$ Perovskites | Formation Energy | 0.81 | 1.18 | 0.49 |
| | Energy Above Hull | 0.82 | 1.16 | 0.28 |

Table 2: Statistical summary of the predictions for three properties collected from two new datasets.

# 7. Application to interfaces

The method was extended to work with interfaces. The layers of interfaces are treated as components of the system.

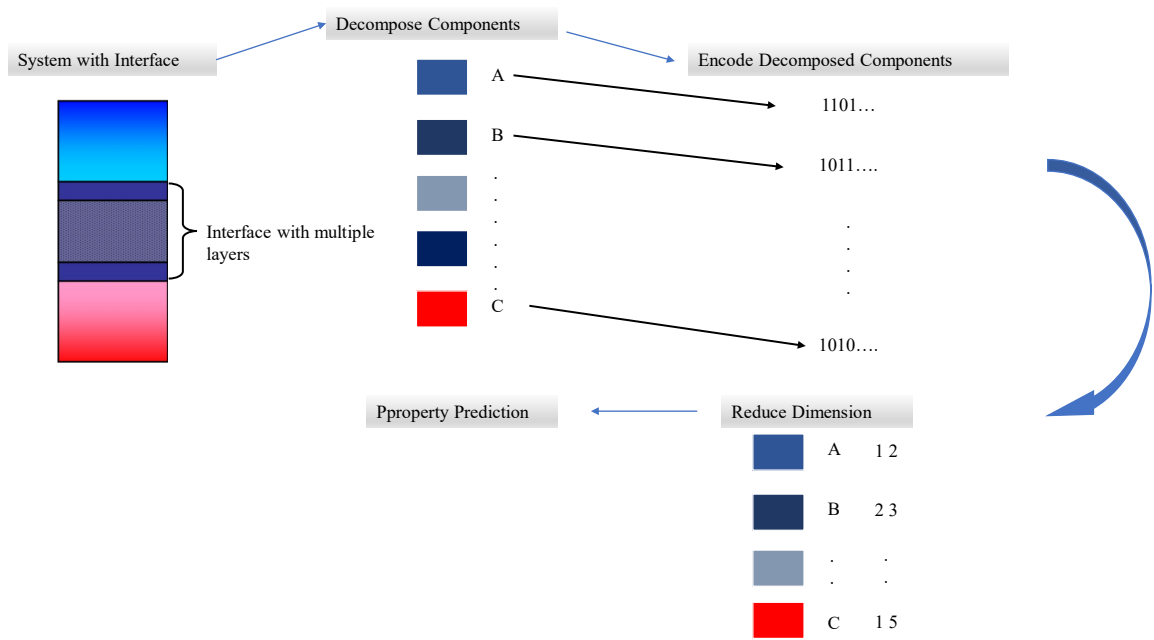

Figure 7: Schematic diagram to show working principal for generating LDFD for interfaces.

Figure 7 shows how LDFD is applied on interfaces. The process to generate LDFD for interfaces is similar to any crystals, here instead of atom fragments, layers are considered as fragments. The remaining steps are the same, converting the fragments to binary and applying logistic PCA.

The interface thermal resistance dataset (ITR) was collected form experimental data in 87 published papers. The ITR dataset contains 1330 data composed of 457 interface samples and 54 materials, including metals, insulators, and semiconductors. The 457 interfaces are defined by their films, interlayers, substrate materials, and experimental conditions. For generating the descriptors for the dataset temperature in conjunction with interface description, as there are systems that are same in terms of components but shows different interfacial property. To generate descriptors for interfaces, the films and substrates were extracted from the dataset and the interlayers were subdivided to layers based on present sublayers. Next, the decomposed components were encoded with one hot encoding and dimension were reduced with logistic PCA. To perform regression the dataset was subdivided into two parts one with zero (0) interlayers and the rest with one or more interlayers.

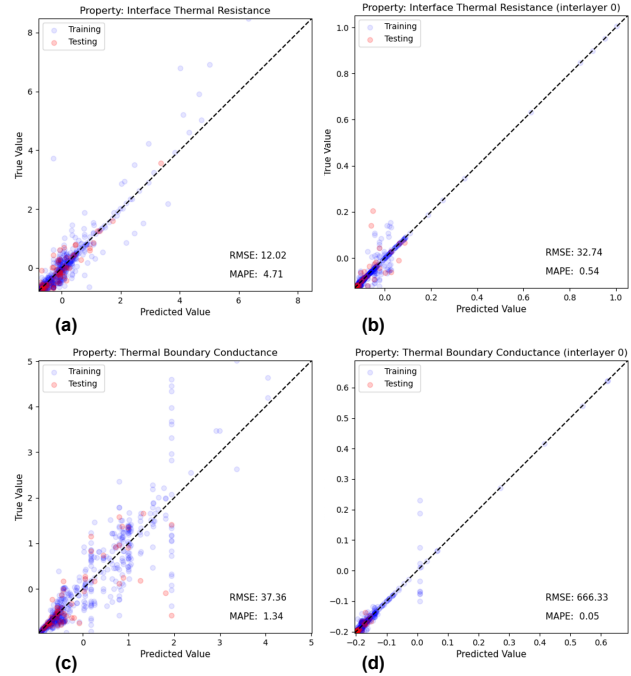

Figure 8: Prediction analysis on ITR dataset. (a-b) Shows Interface thermal resistance prediction for with and without the presence of interlayer. (b-c) Shows Thermal boundary conductance prediction for with and without the presence of interlayer.

The error without interlayer is much higher due to presence of 2$D$ materials with unusually high value for thermal boundary conductance. In both cases, very high value of $R^2$ can be achieved with 10-fold cross validation. Without interlayer $R^2$ value of 0.85 is achieved on average and with interlayer $R^2$ more than 0.95 is achieved. From Figure 6, $RMSE$ is always higher for interlayer 0 materials as these include the 2$D$ materials. Because of very small 2$D$ material data the errors are high. But, even for 2$D$ material systems $MAPE$ is very low indicating provided enough data LDFD should perform well with 2$D$ systems. And with time and advancement in material informatics the unavailability of data should be solved. Similar dataset was studied by Wu *et.al.* [37] they discarded the 2$D$ materials and achieved $R^2$ of 0.95 and $RMSE$ of 10.30 which is very close to the performance by LDFD with lower dimensional descriptors (32 dimensional descriptor vs 11 dimensional descriptor with LDFD descriptor).

| Interlayer | Description | Thermal Boundary Conductance | Interface Thermal Resistance |
|---|---|---|---|
| Without Interlayer | Minimum Value | 0.09 | 0.007 |
| Without Interlayer | 75 percentiles | 190 | 37.04 |
| Without Interlayer | Maximum Value | 135000 | 11111.11 |
| Without Interlayer | Standard Deviation | 10333.14 | 523.67 |
| With Interlayer | Minimum Value | 2.45 | 2.32 |
| With Interlayer | 75 percentiles | 115.3 | 47.77 |
| With Interlayer | Maximum Value | 430 | 407.32 |
| With Interlayer | Standard Deviation | 71.5 | 382.10 |

Table 3: Statistical description of ITR dataset

Very high standard deviation in Thermal boundary conductance causes large $RMSE$ for $2D$ materials.

## 8. Conclusions

Researchers are embracing more and more inclusion of statistical learning and data driven approach in their research and distancing themselves from traditional trial-and-error approach. Data driven statistical approach to understand scientific laws and principles by machine learning is transforming scientific landscape. Though machine learning models are predictive and sometimes not interpretable. ML techniques have provided the researchers with a rational option to circumvent traditional trial-and-error approaches. ML approaches take advantage of available data which accelerates materials discovery. Typical high-throughput DFT calculations do not take advantage of previously calculated data, that's where ML approaches come in and has potential to narrow down search space saving researchers computational cost and time. An average DFT calculation can take thousands of CPU-hours of calculations even when the range of target property is known. In such cases ML models offer an opportunity to leverage previous results for rapid pre-screening of potential materials.

ML methods take full advantage of previously available results, and the presented descriptors are simple and easy to implement and low dimensional to accelerate prediction time. The descriptors only depend on the structural formula which are readily available, so no additional steps are needed for prediction. And offers a prediction rate about $0.1 ms$ per material which should speed up screening process. Several datasets with various properties have been demonstrated to work with this approach and should be capable of discovering new targeted property for new materials. Although the models exhibit excellent predictive power with minor deviations, biased and unbalanced datasets are worst offenders for prediction.

Application of ML approaches usually require sufficient, balanced, and unbiased data for reasonable prediction accuracy. However, available datasets are usually biased toward positive results or lacking sufficient data. There are conflicts within and between datasets of materials which also pose a challenge for predictive performance. To improve large-scale high-throughput computational screening, reliable and accurate data mining approaches need to be developed. In this work, a fragment-based descriptors have been developed to work with inorganic materials. Its performance has been demonstrated with five different datasets and compared with previous work. Larger and balanced datasets should increase its performance.